# Tectonic "short circuit" of sub-horizontal fluid-saturated bodies as a possible mechanism of the earthquake.

Andrei Nechayev                    *Moscow State University, Geographic Department*

*An alternative earthquake mechanism is proposed. The traditional stress mechanism of fracture formation assigned a support role. As a proximate cause of the earthquake the destruction of the roofs of sub-horizontal fluid-saturated bodies (SHFB) is considered. This collapse may occur due to redistribution of fluid pressure within the system of SHFB connected by cracks (tectonic or other nature). It can cause both shifts of rock blocks contributing to seismic shocks and various effects characteristic of foreshocks and aftershocks.*

Earthquake is one of the most formidable and mysterious phenomena of Nature. Its destructive effects are obvious to all as "lying on the surface" but "roots" of terrible events are hidden deep underground. Earthquake is an integral part of rifting as geotectonic process. It can be said that they are genetically related to each other. However, the physical mechanism of the earthquake, as well as its driving force is not determined. Traditional ideas are based entirely on the mechanistic approach and the theory of elasticity: tectonic stresses reaching the tensile strength of rocks leads to the formation of faults, cracks, shifts, etc. [1,2]. However, this mechanism is not able to give a clear physical interpretation of all variety of facts and laws specific to earthquakes and their essential attributes such as foreshocks and aftershocks.

The traditional view is also that the fluids in seismic zones, playing the role of "grease" inside the faults and fractures, may reduce the strength of the mutual blocks friction and the threshold stress leading to the shear-faults and earthquakes.

Under the fluid we mean a liquid or gaseous substance in the earth depth with appropriate pressure and temperature. Fluids thus include a variety of natural substances: magma, water in the form of liquid or steam, as well as combustible gases hydrogen and methane. Volatile fluids are gaseous or can pass it at lower pressure.

The possibility of non-standard effects of deep fluids on the emergence and development of the earthquake was discussed in a number of papers [3,4]. The mechanism of the ascent of the fluid domain in the upper layers of the crust was described in [4]. Excessive fluid pressure thus able to rip open fracture by pushing it upwards until the surface. The authors of [3,4] give numerous evidences of emissions of gases before and during earthquakes, which in some cases appear to be more reliable precursor of the main shock then the problematic measuring of the increased stress of the crust.

In our work, we will proceed from the description of fluid systems given in [5] and use the concept of sub-horizontal fluid-saturated body (SHFB) which define as a closed volume $V$ containing a gaseous fluid (SHFB may contain also the liquid phase but for simplicity we assume the latest unavailable). The vertical size of SHFB (average thickness $h$) is much less than the transverse dimensions: $V \sim hS$, where $S$ is the cross-section of the SHFB. It is important that the rocks filling SHFB should be permeable (porous) to equalize the fluid pressure by the volume. As is known, the state of the fluid being in a supercritical condition with good accuracy satisfies the ideal gas law [6]. Consequently, the fluid pressure *p* can be evaluated as follows:

$$p = \frac{M_f(t)}{V} RT \qquad (1)$$

where $M_f(t)$ is the mass of the fluid which can change over time, *R* is the universal gas constant of the fluid, *T* is the fluid temperature.

SHFB is located at a depth *H*, presumably where there exists an impermeable layer, creating a series of traps. The formation of such SHFB can take place in accordance with the lifting mechanism of sub-vertical fluid domains described in [4]. Excessive fluid pressure expands and extends the micro-crack and the lithostatic pressure difference between the top of the domain and its bottom promotes the fluid upward (in essence, it "floats" as a bubble) until it reaches the barrier, impermeable layer at depth *H*.

The pressure inside the stable SHFB should approach the lithostatic $\rho g H$ where $\rho$ is the average density of the crust. Excessive or insufficient fluid pressure may be offset by the change in SHFB volume due to micro-fractures. Thus, according to [4], we assume that SHFB formation represents the accumulation of fluid when lifting from the upper regions of the mantle or due to chemical transformations of the rocks (e.g. dehydration).

According to the drilling of the famous "Kola super-deep" [7], at depths of up to 5 km were identified distinct fracture zones ranging from 30 to 80 cm, which were repeated every 500-1500 meters. Breeds outside these areas were virtually impenetrable. 9 km below the intensive zones capacity of 10-20 m with high permeability were observed. It can be assumed that a similar pattern exists at large depths where the fluids and water in particular already possess the supercritical status.

For SHFB there must be another important mechanical characteristic: tensile strength (limit of the vertical pressure on the roof) at which the collapse of SHFB occurs. If $P_{str} > \rho g H$, the SHFB failure does not occur even if the vacuum therein. If $P_{str} < \rho g H$ the SHFB preserves its integrity due to the pressure *p* of the fluid, until $\rho g H - p < P_{str}$ or $p > \rho g H - P_{str}$. When the fluid pressure is less than $(\rho g H - P_{str})$, the SHFB roof collapse with possible falling down of the overlying block to a depth *h*. In essence this is a seismic shock. The maximum possible energy of this shock will be, obviously, equal (in case of free fall of the block) to *mgh* which represents the potential energy of a mass *m* located above the SHFB. This power, partly converted into kinetic energy, transforms the energy of braking to the seismic energy.

How do can be done the pressure collapse in SHFB? This will happen if some crack or fracture connect this SHFB with the ground surface or other SHFB, located at substantially less depth with less pressure respectively. Such a "short circuit" can take place in areas of tectonic stress: zones of Subduction, Serpentinization or rifting wherever the crust is subjected to mechanical compression or tension.

SHFB themselves may be nuclei of initiation and propagation of cracks in the direction perpendicular to the direction of compression. Thus, the cracks of different SHFB may overlap, providing them with a "short circuit" and the redistribution of fluid.

Let's consider the intended course of the earthquake on the example of two SHFB located at different levels (Fig. 1).

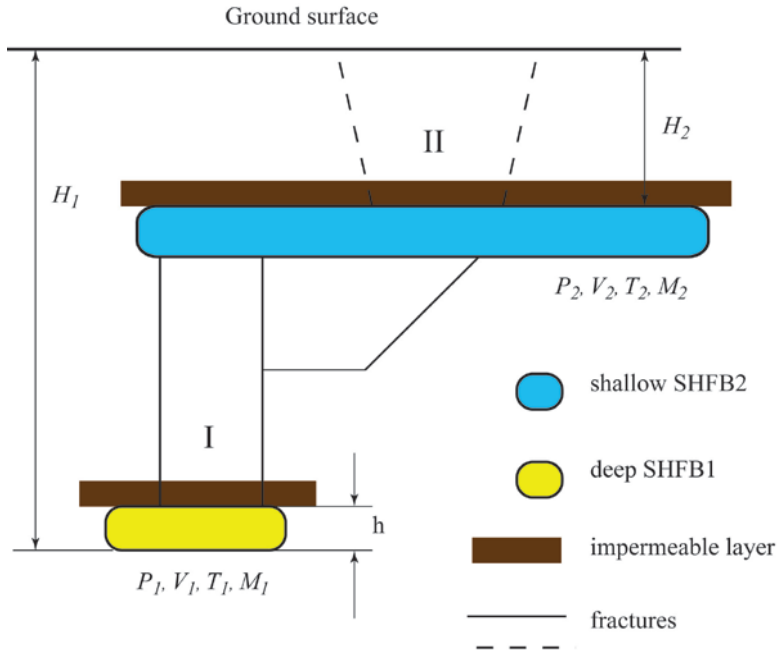

**Fig.1**

*Schematic configuration of two SHFB at different depths. Tectonic fractures connect ("close") them among themselves and (or) to the surface.*

The pressure in each SHFB is equal to the lithostatic one, the other hand, it must satisfy the equation (1). If the crack connects SHFB, there is a redistribution of fluid pressure in accordance with the parameters of each SHFB and hydrodynamic conditions of the total system "SHFB-crack." The average pressure that will be installed in both SHFB is determined by their parameters and above all by its volumes and levels of bedding. Not considering the difference of temperatures and hydrostatic pressure of the fluid in the fracture, the resulting pressure can be estimated by the formula:

$$p_3 \approx \rho g H_2 \frac{V_2}{V_1+V_2} + \rho g H_1 \frac{V_1}{V_1+V_2} \qquad (2)$$

If the decrease of pressure in SHFB1 reach the tensile strength, the block I will collapse (Fig. 1). If the increase of pressure in SHFB2 reach its ultimate strength, the block II will uplift. If the secondary fracture system II (Fig. 1) will contact SHFB2 with the atmosphere due to block II uplift, the drop back of this block may occur as SHFB2 pressure becomes equal to the atmospheric pressure. The above described collapse of the block I can be attributed to the moment of the main shock. The maximum possible energy of this impact will be, obviously, $\rho g(H_1 - H_2)S_I$ where $S_I$ is the block I cross-section. Note that the empirical relation between earthquake magnitude and the linear dimension of the seismic focus [2] is in good agreement with this formula for the energy of the main shock. Indeed, this energy is proportional to the cross section of the block collapsed or to the square of its linear size. Consequently, an increase in the block linear dimension three times (cross-section will be 9 times higher) should increase the magnitude of the earthquake approximately by unit that confirms the data given in [2].

Opening of the cracks in the zone I, the corresponding seismic tremors, the pumping of fluid from SHFB1 to SHFB2 (this process is under pressure in the narrow cracks and it can be accompanied by a drone) reasonably could be regarded as foreshock events. After the collapse of the blocks I and II the injection of fluid (it may be in the supercritical state) within the whole system of cracks occurs and it can activate other SHFB located in the area of "short-circuit", that is in the seismic focus. Reducing the temperature of the fluid at its adiabatic expansion in the zone of lower temperature starts the process of condensation with a corresponding drop in saturated fluid pressure. It can serve as an additional cause of aftershock diverse manifestations. The earthquake process in the model structure (Fig.1) reflects the supposed time dependence of pressure in two SHFB (Fig. 2).

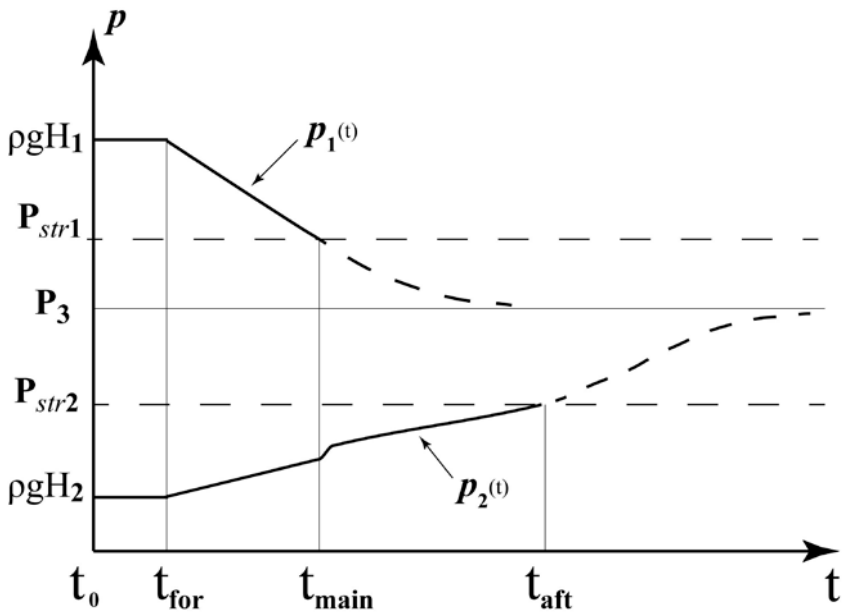

**Fig.2**

*The time dependence of the fluid pressure in two SHFB (Fig. 1), connected by the tectonic crack. $P_{str}$ is the critical fluid pressure at which the collapse of roofs in the corresponding SHFB occurs*

At the time $t_0$ the formation of tectonic fractures in the rock starts. $t_{for}$ is the moment of the "short-circuit" of SHFB with one or more fractures. The fluid pressure redistribution between SHFB starts with the corresponding count of the foreshock events. When pressure in the SHFB1 reaches its tensile strength the block I collapses. This is the moment of the main shock. The pressure in the SHFB2 continues to grow, this process may increase after the collapse of the SHFB1 and corresponding injection of fluid from SHFB1 through the cracks inside SHFB2 (jump on the curve $p_2$). If the pressure inside SHFB2 reaches its ultimate strength (moment $t_{aft}$), there is a reverse fault of the block II and the beginning of aftershock events, which can last as long as the pressure field in the cracks and SHFB come into equilibrium. The duration of aftershocks will depend on the extent and parameters of the fracture along which the pumping of the fluid occurs.

It is clear that the considered system of SHFB is extremely simplistic. The role of SHFB2 can play the system of cracks with the total volume $V_2$ where the fluid coming from SHFB1 is distributed. A corresponding increase in SHFB2 fluid pressure will precede the main shock – the fall down of the block I into SHFB1.

A number of observations and empirical data published in [3,4,8,9] are in good agreement with the above described mechanism of the earthquake. This is primarily an emission of gases (including fuel gases) before, during and after the main shock. These gases being often overheated represent probably fluids passed through the cracks and fissures from the zone of "deep" SHFB1 to the area of "shallow" SHFB2 or directly on the surface of the earth. They could increase the porous fluid pressure at the appropriate levels as it happened during the earthquake in L'Aquila (2009) before the foreshock, and could change the relationship $v_p/v_S$ in the area adjacent to the seismic focus indicating the appearance of fluid-filled dilatancy [8]. Events in L'Aquila may well be interpreted as a consequence of the transfer of fluid from the zone SHFB1 to SHFB2 where the pressure difference first exceeded the threshold strength (foreshock occurred with M = 4), and then exceeded the tensile strength of the roof of SHFB1 with the block I collapse and the corresponding main shock (M = 6.3).

Heavy traffic and pumping of fluids from the lower to the upper horizons in the SHFB "short-circuit" system can lead to all sorts of geophysical phenomena. If cracks through which the fluid is moving do not reach the surface, the increase in pressure in the upper SHFB may be accompanied by "swelling" of the respective geological structures, increasing their size, what was recorded in Japan as a precursor to the earthquake [2]. Ascending flows of fluids in the cracks connecting SHFB with different fluid pressure can cause severe electrification of various rock layers and their subsequent breakdown which can create an electromagnetic pulse with the release of energy in the atmosphere and ionosphere leading to significant disturbances in the Earth's magnetosphere. In some cases these disturbances could also serve as a precursor of the main shock [9].

Thus, the preparation of the earthquake, from our point of view, involves two processes: the accumulation of fluids in the form of SHFB at different depths and the accumulation of tectonic stress on the corresponding blocks. Cracks opening and the "short circuit" of several SHFB is a necessary condition of the earthquake. SHFB destruction is a sufficient condition. Earthquake power depends on what SHFB on what depths are involved in the process of fluid redistribution. The magnitude of the main shock of the earthquake should be determined by the parameters $H, h, S, T, R$ of the deepest destroyed SHFB and the dimensions of the blocks fallen down.

Fault zone provide the "shorting" of SHFB making it more probable. A zone of high seismic activity must include areas where tectonic (or otherwise) cracking is combined with a relatively high rate of generation and accumulation of fluid in the presence of a sufficient number of impermeable layers. The "calm" can be explained by the accumulation of fluid in the SHFB previously devastated by the earthquake. Perhaps the "calm" occurs when the accumulation of fluid goes in the largest SHFB and need macro-fissure to "close" this SHFB and the atmosphere. In this case the pressure drop in the SHFB and the corresponding seismic shock will be maximized.

Thus, we do not reject the mechanism of cracks formation by self-organization at a certain critical condition. We argue that this mechanism is necessary, but not sufficient. He can drive the mechanism of hydrodynamic "short circuit" of the fluid-saturated sub-horizontal bodies leading to the corresponding collapses and seismic shocks and release of seismic energy. Fractal block structure can exist on its own. It forms a network of cracks that give way to the fluid transfer, and they, in turn, are redistributed between SHFB connected by cracks. The result is a new field of the porous fluid pressure. In the places, where this pressure exceeds the tensile strength, the seismic shocks occur.

As the conclusions of the work we present a brief list of obvious advantages of the above-described mechanism of "short-circuit" of sub-horizontal fluid-saturated bodies.

1. This mechanism provides the necessary free space for the seismic shock and all kinds of possible block motion: collapse (SHFB roof destruction due to the overpressure from outside), block uplift (overpressure from inside), oblique movements along fractures and crack planes intersecting SHFB.

2. The proposed mechanism allows for a simple physical way to estimate the energy of the seismic shock.

3. Easily and naturally explains the physical nature of the foreshocks and aftershocks, as well as acoustic and emission effects before, during and after the main shock.

4. Mechanism can account for both "hard" and "soft" (quiet) class of the earthquake.

5. Mechanism explains earthquake without disturbing the Earth's surface (collapse occur within the system with the destruction of some SHFB "cavities" and extension of other at constant total volume of seismic focus area). It explains also the deep-earthquake when the formation of tectonic fractures becomes unlikely because of rocks plasticity but "short circuit" between deep SHFB can occur due to dehydration and hydro-fracturing.

6. This mechanism should be well described by the Gutenberg-Richter power law since it corresponds to the classical model of forest fires [10]. The cross section of SHFB is analogous to the size of forest cluster, the accumulation of fluid corresponds to the growth and reproduction of trees, and the tectonic crack, interconnecting SHFB, is an analogue of lightning randomly igniting forest.